\documentclass[authoryear,preprint,review,12pt]{elsarticle}

\usepackage{amssymb}

\usepackage{natbib}
\setcitestyle{square,numbers,sort&compress,comma}
\usepackage[flushleft]{threeparttable}
\usepackage[labelsep=period]{caption}
\usepackage{amsmath}
\usepackage{siunitx}
\journal{-}

\begin{document}

\begin{frontmatter}



\title{ Enhanced fracture toughness in ceramic superlattice thin films: on the role of coherency stresses and misfit dislocations}


\author[1]{Antonia Wagner\corref{cor1}}
\author[2]{David Holec}
\author[1]{Paul Heinz Mayrhofer}
\author[1]{Matthias Bartosik}

\address[1]{Institute of Materials Science and Technology, TU Wien}
\address[2]{Department of Materials Science, Montanuniversit\"at Leoben}
\cortext[cor1]{Corresponding author}

\begin{abstract}
Superlattice (SL) thin films composed of refractory ceramics unite extremely high hardness and enhanced fracture toughness; a material combination often being mutually exclusive. While the hardness enhancement obtained when two materials form a superlattice is well described by existing models based on dislocation mobility, the underlying mechanisms behind the increase in fracture toughness are yet to be unraveled. 
Here we provide a model based on linear elasticity theory to predict the fracture toughness enhancement in (semi-)epitaxial nanolayers due to coherency stresses and formation of misfit dislocations. We exemplarily study a superlattice structure composed of two cubic transition metal nitrides (TiN, CrN) on a MgO (100) single-crystal substrate. Minimization of the overall strain energy, each time a new layer is added on the nanolayered stack, allows estimating the density of misfit dislocations formed at the interfaces. The evolving coherency stresses, which are partly relaxed by the misfit dislocations, are then used to calculate the apparent fracture toughness of respective SL architectures by applying the weight function method. 
The results show that the critical stress intensity increases steeply with increasing bilayer period for very thin (essentially dislocation-free) SLs, before the ${K_\mathrm{IC}}$ values decline more gently along with the formation of misfit dislocations. The characteristic ${K_\mathrm{IC}}$ vs. bilayer-period-dependence nicely matches experimental trends. Importantly, all critical stress intensity values of the superlattice films clearly exceed the intrinsic fracture toughness of the constituting layer materials, evincing the importance of coherency stresses for increasing the crack growth resistance.
\end{abstract}


\end{frontmatter}


\section{Introduction}
\label{sec:int}
A careful microstructural design has proven to be an effective strategy for enhancing the performance of ceramic thin films. Particularly the alternating deposition of two or more different layer materials with a periodicty in the nanometer range allows for tailored properties that exceed the inherent properties of the multilayers monolithic constituents~\cite{Holleck1995,Stueber2009}.\\
Helmersson et al.~\cite{Helmersson1987} showed that hardness values of 5560 HV (about \SI{55}{GPa}) can be obtained by growing TiN and VN in a superlattice architecture, which is a more than 100\% increase as compared with single-phase TiN or VN. The multilayer structure comprised nanometer-thin iso-structured TiN and VN layers, which were epitaxially grown on a single-crystalline MgO substrate by means of physical vapour deposition. For a better understanding of the mechanisms behind this effect, Chu and Barnett~\cite{Chu1995} proposed a model based on dislocation glide within the individual layers and across the interfaces. As revealed by recent micro-mechanical tests on nitride-based superlattice thin films~\cite{Hahn2016a,Buchinger2019}, besides the hardness also the fracture toughness is enhanced in the multilayered nanostructure; two material properties which are often mutually exclusive~\cite{Ritchie2011} but of high relevance for inherently brittle refractory ceramics. Both, hardness and fracture toughness, exhibit a strong dependence on the bilayer period of the superlattice with its peak at a few nanometers. Unlike the hardness enhancement, the mechanisms behind the superlattice effect in terms of the fracture toughness are far less understood. Possible mechanisms are for instance: coherency stresses and misfit dislocation arrays, elastic mismatch between the layer materials, and the change of the bonding characteristics with decreasing layer thicknesses.\\
Coherency stresses can reach high values in the order of tens of GPa in the absence of misfit dislocations in epitaxial nanolayers. A fully coherent layer growth, however, is rather unlikely for lattice mismatched heterostructures, since the formation of misfit dislocations becomes energetically favourable when a certain layer thickness is exceeded. This critical thickness concept was first proposed for a single coherently grown layer by Frank and van der Merwe~\cite{FrankMerwe1949} based on an energy minimization of strain energy in the layer and the energy due to the local dislocation strain field. Some years later Matthews and Blakeslee~\cite{Matthews1974} derived a criterion for misfit accommodation based on mechanical equilibrium theory considering the force exerted by the misfit strain and the tension in the dislocation line. These models have been extended by numerous researchers, e.g. regarding elastically anisotropic material behavior, interaction between dislocations, the effect of introducing dislocations sequentially and asymmetric misfit strain~\cite{Willis1990,Atkinson1992,Holec2007,Andersen2017}. Still, there appears to be little research on dislocation density in multilayer structures and the literature available on this topic only considers equal dislocation density in all layers or even treats the superlattice as an alloy regarding its lattice parameter~\cite{ Matthews1974,Hirth1990,VanDeLeur1988}. \\
The objective of the present work is to elaborate the role of coherency stresses and misfit dislocations for the enhanced fracture toughness of ceramic superlattices. We will develop an algorithm to determine the dislocation density and the resulting residual stress state of a (semi-)coherently grown multilayer. Based on the stress state, we will then calculate the fracture toughness of the system under consideration by adapting the weight function method proposed by Bueckner et al.~\cite{Bueckner1970}. 

\section{Method}
\label{sec:method}
\subsection{Estimation of the dislocation density in a single layer}

\begin{figure}[htb]
\centering
\includegraphics[scale=0.37]{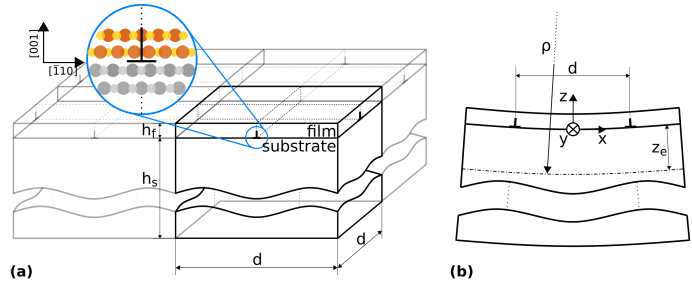}
\caption[Three-dimensional unit cell of the film-substrate system with a network of orthogonal edge dislocation arrays (a), Euler-Bernoulli beam theory model with deposition-induced curvature (b)]{Three-dimensional unit cell of the film-substrate system with a network of orthogonal edge dislocation arrays (a), Euler-Bernoulli beam theory model with deposition-induced curvature (b)}
\label{pic:system_def}
\end{figure}

First, we study the simplest configuration of a single layer deposited on a substrate, see Fig.~\ref{pic:system_def}a. The film grows coherently up to a critical thickness, $h_\mathrm{crit}$, at which part of the lattice mismatch between film and substrate material starts being accommodated by misfit dislocations. Based on experimental high resolution transmission electron microscopy analyses of cubic transition metal nitride thin films, e.g. by Hultman et al.~\cite{Hultman1994}, we consider a network of orthogonal edge-type dislocation arrays as depicted in Fig.~\ref{pic:system_def}a. We presume, that the orthogonal dislocations form simultaneously. Hence, a biaxial stress state is considered at any point of the layer addition. With growing layer thickness the mean distance between dislocations $d$ is reduced, i.e. the areal dislocation density $Q$, with unit $\mathrm{{cm}^{-1}}$, increases.  We calculate the density of dislocations at the interface by minimizing the overall elastic energy in the system with respect to $Q$. The elastic energy per interface area $\frac{E(Q)}{A}$ of the system is given by the sum of the energy due to the mean strain in the layer $\frac{E_\mathrm{strain}(Q)}{A}$ and the dislocation energy resulting from the local dislocation stress field $\frac{E_\mathrm{disloc}(Q)}{A}$:
\begin{equation}
\frac{E(Q)}{A}= \frac{E_\mathrm{strain}(Q)}{A}+\frac{E_\mathrm{disloc}(Q)}{A}.
\label{equ:energy}
\end{equation}

For evaluating $\frac{E_\mathrm{strain}(Q)}{A}$, the stress state of the system as a function of $Q$ has to be defined. Within this treatment the dislocation nucleation process is disregarded.\\
We consider the film to be much thinner than the substrate and expect the strain relaxation by substrate bending to be of minor relevance. Nevertheless, we include the induced curvature, since it will enable us to validate our results with substrate curvature measurements of a growing superlattice in potential follow-up studies. 
In accordance with Euler-Bernoulli beam theory, see Fig.~\ref{pic:system_def}b, the relation between the normal strain $\varepsilon_{x} (z)$, the radius of curvature $\rho$ and the position of the axis with zero strain $z_{e}$ can be written as
\begin{equation}
\varepsilon_{x}(z)=\frac{z-z_{e}}{\rho}.
\end{equation}
In order to estimate the two unknowns $\rho$ and $z_{e}$, we define the force and moment balance of the substrate-film system. Considering a biaxial stress state parallel to the interface we get:
\begin{equation}
\label{equ:force_balance_single}
F_\mathrm{s}+F_\mathrm{f}=\int_{-h_\mathrm{s}}^{0} \! \overline{M_\mathrm{s}} \Bigl(\frac{z-z_{e}}{\rho}\Bigr) \, \mathrm{d}z+\int_{0}^{h_\mathrm{f}} \!\overline{M_\mathrm{f}} \Bigl(\frac{z-z_{e}}{\rho}+\varepsilon_\mathrm{m}\Bigr)\, \mathrm{d}z
=
0,
\end{equation}
\begin{equation}
\label{equ:moment_balance_single}
M_\mathrm{s}+M_\mathrm{f}=\int_{-h_\mathrm{s}}^{0} \! \overline{M_\mathrm{s}} \Bigl(\frac{z-z_{e}}{\rho}\Bigr) z\, \mathrm{d}z+\int_{0}^{h_\mathrm{f}} \!\overline{M_\mathrm{f}} \Bigl(\frac{z-z_{e}}{\rho}+\varepsilon_\mathrm{m}\Bigr) z\, \mathrm{d}z
=
0,
\end{equation}
where $F_\mathrm{s}$ and $F_\mathrm{f}$ represent the resulting force in the substrate and the layer, respectively, and $M_\mathrm{s}$ and $M_\mathrm{f}$ the bending moment in the substrate and the film with reference to $z=0$ for a system with unit thickness in $y$-direction. 
The mean strain $\varepsilon_\mathrm{m}$ due to relaxed lattice mismatch depends on the lattice parameters of the constituents, the component of the Burgers vector parallel to the interface and the dislocation density:
\begin{equation}
\label{equ:eps_m_single}
\varepsilon_\mathrm{m}=\frac{a_\mathrm{s}[1-\mathrm{sgn}(a_\mathrm{s}-a_\mathrm{f})b_xQ]-a_\mathrm{f}}{a_\mathrm{f}}.
\end{equation}
$\overline{M_\mathrm{s}}$ and $\overline{M_\mathrm{f}}$ denote the biaxial moduli of the substrate and the layer.
For cubic crystal symmetry the biaxial modulus of a \{001\} plane is: 
\begin{equation}
\label{equ:biaxial_modulus}
\overline{M}=c_{11}+c_{12}-\frac{2 c_{12}^2}{c_{11}},
\end{equation}
with
\begin{equation}
c_{11}=C_{iiii},
\qquad
c_{12}=C_{iijj},
\qquad
c_{44}=C_{ijij},
\qquad
i,j=1,2,3.
\end{equation}

By substituting Eqs.~\ref{equ:eps_m_single} and \ref{equ:biaxial_modulus} into Eqs.~\ref{equ:force_balance_single} and \ref{equ:moment_balance_single} we determine the axis of zero strain and the curvature of the film-substrate system as a function of dislocation density ${Q}$. The strain energy per interface area is then evaluated by:
\begin{equation}
\frac{E_\mathrm{strain}(Q)}{A}=\frac{1}{2} \int_{-h_\mathrm{s}}^{0} \! \sigma_\mathrm{s} \Bigl(\frac{z-z_e}{\rho}\Bigr)\, \mathrm{d}z+\frac{1}{2} \int_{0}^{h_\mathrm{f}} \! \sigma_\mathrm{f} \Bigl(\frac{z-z_e}{\rho}+\varepsilon_\mathrm{m}\Bigr)\, \mathrm{d}z,
\label{equ:strain_energy}
\end{equation}
with the biaxial stress states in the substrate and the film being:
\begin{equation}
\sigma_\mathrm{s}=\overline{M_\mathrm{s}}\Bigl(\frac{z-z_e}{\rho}\Bigr)\hspace{20pt}
(-h_\mathrm{s} \leq z \leq 0),
\label{equ:stress_sub}
\end{equation}
\begin{equation}
\sigma_\mathrm{f}=\overline{M_\mathrm{f}}\Bigl(\frac{z-z_e}{\rho}+\varepsilon_\mathrm{m}\Bigr)
\hspace{20pt}
(0 \leq z \leq h_f).
\label{equ:stress_film}
\end{equation}
Low dislocation densities justify the assumption of non-interacting dislocations within an array. It follows that the dislocation energy per interface area can be calculated by the line energy of a single dislocation mutliplied by the dislocation density. In the region of the dislocation core with a radius ${r_\mathrm{c}}$ the theory of linear elasticity breaks down. We therefore divide the dislocation line energy into two parts, namely the nonelastic core energy per dislocation line length, $\frac{E_\mathrm{core}}{L}$, and the linear elastic dislocation line energy outside the core, $\frac{E_\mathrm{disloc,le}}{L}$. 
Hence, Eq.~\ref{equ:energy} becomes:
\begin{equation}
\frac{E(Q)}{A}= \frac{E_\mathrm{strain}(Q)}{A}+\Bigl(\frac{E_\mathrm{disloc,le}}{L}+\frac{E_\mathrm{core}}{L}\Bigr) Q.
\label{equ:energy_total}
\end{equation}
For the sake of simplicity, the linear elastic part of the dislocation line energy is assessed by considering an infinite medium with elastic properties estimated by the arithmetic average of the adjacent constituents. According to Foreman~\cite{Foreman1955} the elastic energy induced by a dislocation is:
\begin{equation}
\frac{E_\mathrm{disloc,le}}{L}=\frac{K b^2}{4\pi}ln\Big(\frac{R}{r_\mathrm{c}}\Bigl),
\label{equ:disloc_energy}
\end{equation}
where $K$ is the energy factor, being dependent on the material's anisotropy and the type of dislocation. For an edge dislocation in a cubic crystal $K$ is given by
\begin{equation}
K=(c_{11}+c_{12})\left[\frac{c_{44}(c_{11}-c_{12})}{c_{11}(c_{11}+c_{12}+2c_{44}}\right]^{\frac{1}{2}}.
\end{equation}
$R$ represents the outer cut-off radius, which we set equal to the thickness of the deposited layer $h_\mathrm{f}$. The dislocation core radius $r_\mathrm{c}$ is taken to be equal to $b$. For transition metal nitrides the dislocation core line energy typically lies between 1\texttt{-}\SI{2}{eV/\AA} hence we investigate the influence of $\frac{E_\mathrm{core}}{L}$ within these boundaries. Even though the impact of the core energy is not negligible especially for small layer thicknesses, the qualitative behavior is rather unaffected.\\
The equilibrium configuration of the system is evaluated by minimizing Eq.~\ref{equ:energy_total} with respect to ${Q}$, a concept first proposed by Frank and van der Merwe~\cite{FrankMerwe1949}. A negative value of ${(a_\mathrm{s}-a_\mathrm{f})}$ indicates that the lattice mismatch strain is reduced by removing a half plane of width $b_x$ at a distance of ${d=\frac{1}{Q}}$.\\

\subsection{Algorithm for the dislocation density in a multilayer}
\label{multilayer_algorithm}
\begin{figure}[htb]
\centering
\includegraphics[scale=0.47]{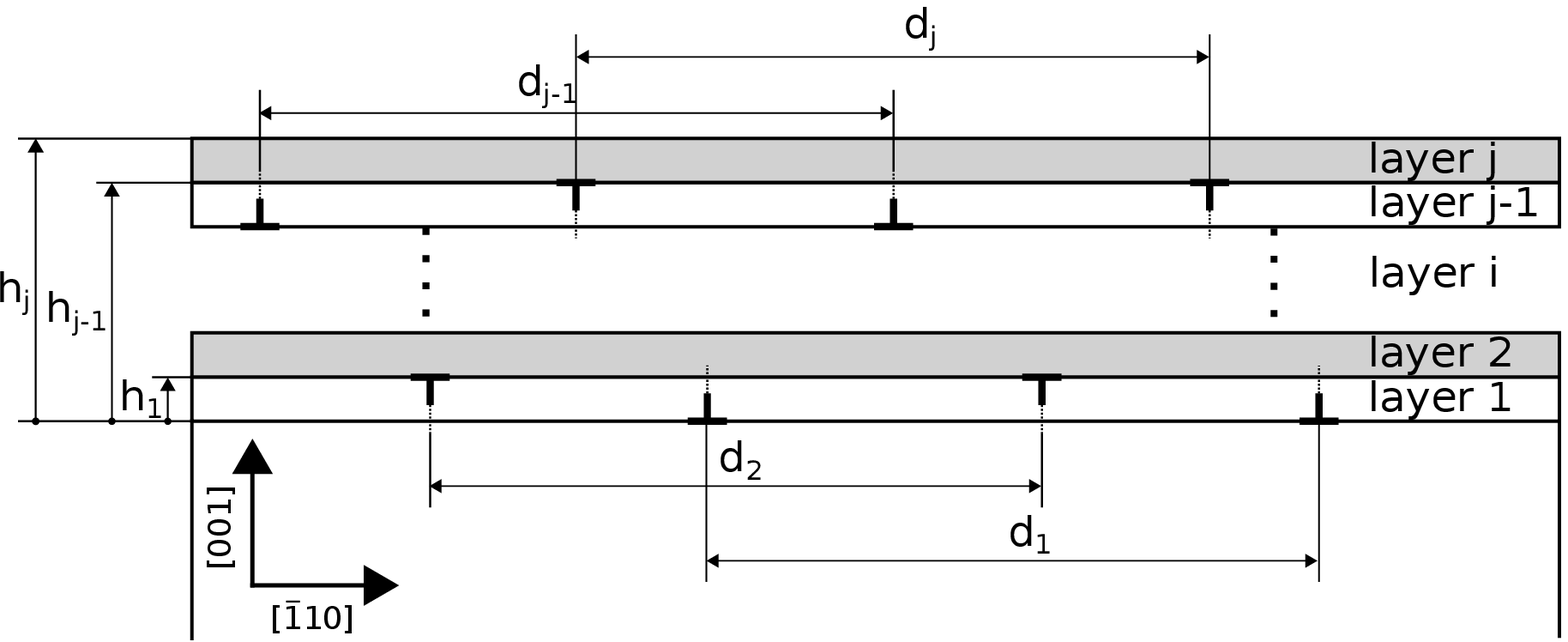}
\caption[a]{Schematic of the considered dislocation distribution in a superlattice of two different transition metal nitrides with Burgers vectors in the $\left[\overline{1}10\right]$ direction and dislocations of opposite sign in subsequently deposited layers. Both configurations, a perfectly sharp interface and a finite interface thickness with composition modulation, are studied.}
\label{pic:scheme_multi}
\end{figure}

We consider a layerwise deposition of the superlattice. Hence, the equilibrium state after each layer addition plays a decisive role for the formation of dislocations in the subsequent layers. Figure~\ref{pic:scheme_multi} schematically depicts a superlattice with dislocations at the interfaces between two different transition metal nitrides with the direction of the Burgers vector parallel to the $\left(\overline{1}10\right)$ direction and a value of $b_{x,j}=a_j\frac{1}{\sqrt{2}}$. The alternating stress fields in a superlattice lead to misfit dislocations of opposite sign in consecutive layers.\\
The force and moment balances have to be fulfilled at each step of the layerwise assembling of the multilayer. Consequently, according to the superposition principle, we can define the balances considering the change of forces and moments due to the $j$-th layer addition:
\begin{equation}
\label{equ:force_balance_multi}
\begin{aligned}
\Delta F_\mathrm{s}+\sum \limits_{i=1}^{j} \Delta F_i=
&\int_{-h_\mathrm{s}}^{0} \! \overline{M_\mathrm{s}} \Bigl(\frac{z-z_e^{(j)}}{\rho^{(j)}}\Bigr) \, \mathrm{d}z\\
&+\sum \limits_{i=1}^{j-1} \int_{h_{i-1}}^{h_i} \!\overline{M_i} \Bigl(\frac{z-z_e^{(j)}}{\rho^{(j)}}\Bigr)\, \mathrm{d}z \\ 
&+\int_{h_{j-1}}^{h_j} \!\overline{M_j} \Bigl(\sum \limits_{k=1}^{j} \frac{z-z_e^{(k)}}{\rho^{(k)}}+\varepsilon_{\mathrm{m},j}\Bigr)\, \mathrm{d}z
=
0,
\end{aligned}
\end{equation}
\begin{equation}
\label{equ:moment_balance_multi}
\begin{aligned}
\Delta M_\mathrm{s}+\sum \limits_{i=1}^{j} \Delta M_i=
&\int_{-h_\mathrm{s}}^{0} \! \overline{M_\mathrm{s}} \Bigl(\frac{z-z_e^{(j)}}{\rho^{(j)}}\Bigr) z\, \mathrm{d}z\\
&+\sum \limits_{i=1}^{j-1} \int_{h_{i-1}}^{h_i} \!\overline{M_i} \Bigl(\frac{z-z_e^{(j)}}{\rho^{(j)}}\Bigr) z\, \mathrm{d}z\\
&+\int_{h_{j-1}}^{h_j} \!\overline{M_j} \Bigl(\sum \limits_{k=1}^{j} \frac{z-z_e^{(k)}}{\rho^{(k)}}+\varepsilon_{\mathrm{m},j}\Bigr) z\, \mathrm{d}z
=
0.
\end{aligned}
\end{equation}

Herein, the superscript $(j)$ denotes the induced curvature/axis of zero strain due to the adding of the topmost layer and $h_j$ denotes the distance between the 	substrate-film interface and the top of layer $j$, see Fig.~\ref{pic:scheme_multi}, with $h_{0}=0$.

The mean misfit strain in the currently added layer is defined as
\begin{equation}
\varepsilon_{\mathrm{m},j}=\frac{\hat{a}_{j-1}\left[1-\mathrm{sgn}(\hat{a}_{j-1}-a_j)b_{x,j} Q_j\right]-a_j}{a_j},
\end{equation}
with $\hat{a}_{j-1}$ being the lattice parameter of the previously deposited layer without curvature considering the formed dislocations:
\begin{equation}
\hat{a}_{j-1}=\hat{a}_{j-2}\left[1-\mathrm{sgn}(\hat{a}_{j-2}-a_{j-1})b_{x,j-1}Q_{j-1}\right],
\end{equation}
with $\hat{a}_{0}=a_\mathrm{s}$.\\
Similar to the deposition of a single layer, the axis of zero strain $z_e^{(j)}$ and the induced curvature $\rho^{(j)}$ can be calculated from Eq.~\ref{equ:force_balance_multi} and Eq.~\ref{equ:moment_balance_multi} as a function of dislocation density $Q_j$.\\
The strain state of the system after $j$ layer depositions results from the superposition of the induced bending of all layers and the relaxed misfit strain:
\begin{equation}
\label{equ:sub_stress_superpos}
\varepsilon_{x,\mathrm{s}}(z)=\sum \limits_{k=1}^{j} \frac{z-z_e^{(k)}}{\rho^{(k)}}
\hspace{20pt}
(h_\mathrm{s} \leq z \leq 0),
\end{equation}
\begin{equation}
\label{equ:layer_stress_superpos}
\varepsilon_{x,i}(z)=\sum \limits_{k=1}^{j} \frac{z-z_e^{(k)}}{\rho^{(k)}}+\varepsilon_{m,i}
\hspace{20pt}
(i=1,...n)
\hspace{20pt}
(h_{i-1} \leq z \leq h_i).
\end{equation}
The biaxial stress state in the layers and the substrate after the deposition of $n$ layers is:
\begin{equation}
\label{equ:sub_stress_superpos}
\sigma_{x,\mathrm{s}}(z)=\overline{M_\mathrm{s}} \varepsilon_{x,\mathrm{s}}(z)
\hspace{20pt}
(h_\mathrm{s} \leq z \leq 0),
\end{equation}
\begin{equation}
\label{equ:layer_stress_superpos}
\sigma_{x,i}(z)=\overline{M_i} \varepsilon_{x,i}(z)
\hspace{20pt}
(i=1,...n)
\hspace{20pt}
(h_{i-1} \leq z \leq h_i).
\end{equation}

The strain energy per interface area as a function of the dislocation energy of the topmost layer is then given by:
\begin{equation}
\frac{E_\mathrm{strain}(Q_j)}{A}=\frac{1}{2} \int_{-h_\mathrm{s}}^{0} \! \sigma_{x,\mathrm{s}} \varepsilon_{x,\mathrm{s}} \mathrm{d}z+\frac{1}{2} \sum \limits_{i=1}^{j} \int_{h_{i-1}}^{h_i} \! \sigma_{x,i} \varepsilon_{x,i} \mathrm{d}z.
\label{equ:strain_energy_multi}
\end{equation}
The total energy becomes:
\begin{equation}
\frac{E(Q_j)}{A}= \frac{E_\mathrm{strain}(Q_j)}{A}+\Bigl(\frac{E_\mathrm{disloc,le}}{L}+\frac{E_\mathrm{core}}{L}\Bigr) Q_j+\frac{E_\mathrm{disloc,prev}}{A},
\label{equ:energy_total_multi}
\end{equation}
where $\frac{E_\mathrm{disloc,prev}}{A}$ represents the dislocation line energy due to the dislocations formed in the previously deposited layers, not being a function of $Q_j$. Minimizing the total energy with respect to $Q_j$ after each layer deposition, we get the dislocation density in all layers and, hence, the stress/strain distribution after $j$ layer depositions.\\
The just described algorithm can be vastly simplified by neglecting the induced curvature and strain in the substrate, i.e. considering the substrate to be thick enough to be idealized as a half space. This reduces Eq.~\ref{equ:strain_energy_multi} to
\begin{equation}
\frac{E^*_\mathrm{strain}(Q_j)}{A}=\frac{1}{2} \sum \limits_{i=1}^{j} \int_{h_{i-1}}^{h_i} \! \sigma_{x,i} \varepsilon_{\mathrm{m},i} \mathrm{d}z,
\label{equ:strain_energy_multi_simple}
\end{equation}
with ${\sigma_{x,i}}$ and ${\varepsilon_{\mathrm{m},i}}$ being constant between ${h_{i-1}}$ and ${h_i}$ and only the last summand depending on ${Q_j}$. The contribution to the total elastic energy per interface area ${\frac{E^*(Q_j)}{A}}$ being a function of ${Q_j}$ is thus reduced to:

\begin{equation}
\label{equ:energy_total_multi_simple}
\begin{aligned}
\frac{E^*(Q_j)}{A}=
&\frac{1}{2}(h_j-h_{j-1})\overline{M_j} \Bigl(\frac{\hat{a}_{j-1}(1-\mathrm{sgn}(\hat{a}_{j-1}-a_j)b_{x,j} Q_j)-a_j}{a_j}\Bigr)^2\\
&+\Bigl(\frac{E_\mathrm{disloc,le}}{L}+\frac{E_\mathrm{core}}{L}\Bigr) Q_j.
\end{aligned}
\end{equation}
By differentiating with respect to ${Q_j}$ and solving for $\frac{\partial}{\partial Q_j} =0$, we obtain the dislocation density that minimizes the elastic energy per interface area.
The dislocation density ${Q_j}$ can then be estimated by
\begin{equation}
Q_j= \frac{|a_{j}-\hat{a}_{j-1}|}{\hat{a}_{j-1} b_{x,j}}-\Bigl(\frac{E_\mathrm{disloc,le}}{L}+\frac{E_\mathrm{core}}{L}\Bigr)\frac{a_j^2}{\overline{M}_j (h_j-h_{j-1}) \hat{a}_{j-1}^2 b_{x,j}^2},
\label{equ:disloc_density_simple}
\end{equation}
with
\begin{equation}
\hat{a}_{j-1}= a_s\prod_{n=1}^{j-1}\Bigl(1-\mathrm{sgn}(\hat{a}_{n-1}-a_{n})b_{x,n} Q_n\Bigr).
\label{equ:a_hat_simple}
\end{equation}

\begin{figure}[htb]
\centering
\includegraphics[scale=0.35]{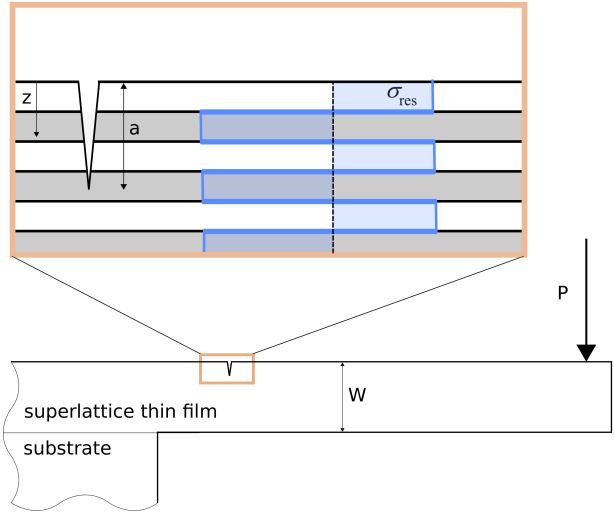}
\caption[Weight function method] {Schematic drawing of the considered micro-cantilever of free-standing film material. The detail depicts the configuration the weight function is applied to.}
\label{pic:weight_function}
\end{figure}

It should be noted that negative values resulting from Eq.~\ref{equ:disloc_density_simple} have to be set to zero and a negative or positive sign of $(a_{j}-\hat{a}_{j-1})$ indicates whether a half plane of $b_{x,j}$ is removed or added, respectively.\\
Perfectly sharp interfaces are often not achievable by physical vapour deposition. Therefore, we further study the effect of a finite interface thickness ${h_\mathrm{interf}=m\overline{a}}$ with ${\overline{a}}$ being the arithmetic mean value of the lattice parameter inherent to the adjacent layer materials and $m$ being an integer number. The interface thickness is assumed to be independent of the bilayer period. A stepwise composition change is considered. Based on the interface thickness of $m$ atomic layers, it is reasonable to define $m+1$ composition steps. Dislocations are assumed to form at the $z$-position where the lattice misfit strain changes its sign, being at the center of the interface for the studied configurations. \\
We intend to simulate the experimental evaluation of the fracture toughness of a superlattice performed on a micro-cantilever of free-standing film material. Therefore, we "remove" the substrate material in our model and determine the new equilibrium state, while keeping the estimated dislocation densities constant. The following assumptions are made at this point: The resulting stress state corresponds to a film entirely free of substrate material and, hence, slightly deviates from the actual boundary condition of a micro-cantilever, see Fig.~\ref{pic:weight_function}. Furthermore, we consider a pure bending stress distribution induced by the load $P$ and neglect the induced shear stresses. 

\subsection{Apparent fracture toughness}
From the method discussed in Sec.~\ref{multilayer_algorithm}, we get an estimation for the residual stress state of the superlattice system after the entire deposition process. In this section, we aim to investigate the influence of the residual stress state on the crack growth resistance of the system. The stress intensity factor present in a cracked material is a superposition of the stress intensity due to external loading and the stress intensity due to the residual stress state. We now consider the contribution of the residual stresses as an alteration of the maximum bearable stress intensity of the system and can define this apparent fracture toughness ${K_\mathrm{app}}$ as:
\begin{equation}
\label{equ:stress_intensity}
K_\mathrm{app}=K_\mathrm{IC}-K_\mathrm{res},
\end{equation}
with ${K_\mathrm{IC}}$ being the inherent Mode I critical stress intensity factor and ${K_\mathrm{res}}$ the stress intensity factor due to residual stresses.\\
The contribution of the residual stress state to the apparent fracture toughness is assessed by means of the weight function theory first proposed by Bueckner~\cite{Bueckner1970}. It allows to estimate the stress intensity factor as a function of crack length $a$ for a given geometry with arbitrary stress distribution.

The stress intensity factor associated with the residual stress state $\sigma_\mathrm{res}$ is given by
\begin{equation}
\label{equ:res_stress_intensity}
K_\mathrm{res}(a)=\int_{0}^{a} h(z,a) \sigma_\mathrm{res}(z) \mathrm{d}z,
\end{equation}
where $z$ is the distance along the crack from the top surface and $h(z,a)$ is the weight function. According to our experimental setup we apply a weight function suitable for an edge cracked bar of thickness ($W$) derived by Fett~\cite{Fett1990}:
\begin{equation}
h(z,a)=\sqrt{\frac{2}{\pi a}}\frac{1}{\sqrt{1-\frac{z}{a}}(1-\frac{a}{W})^\frac{3}{2}}\left[\Big(1-\frac{a}{W}\Big)^\frac{3}{2}+\sum A_{\nu\mu}\Big(1-\frac{z}{a}\Big)^{\nu+1}\Big(\frac{a}{W}\Big)^\mu\right]
\end{equation}
The values of the coefficients $A_{\nu\mu}$ are given in \cite{Fett1990}.

\subsection{Sample superlattice configuration}
We consider elastically anisotropic layer materials on an elastically anisotropic finite substrate, all constituents having a face-centered-cubic structure. The elastic properties of the layer materials investigated within this study are extracted from Density Functional Theory studies~\cite{Buchinger2019,Zhou2013} and summarized in Table~\ref{tab:elast_prop}. 
We assume a (semi-)coherent interface between the MgO substrate and the first layer. The substrate thickness is defined to be \SI{500}{\micro m} for all calculations. If not stated otherwise, a perfectly sharp interface is considered. All sublayers are of equal thickness and the bilayer period is represented by ${\Lambda}$.

\begin{table*}[htb]
\centering
\caption{Properties of the film and substrate materials}
\begin{threeparttable}
\begin{tabular}{p{2.2cm}p{1.5cm}p{1.5cm}p{1.5cm}p{1.5cm}p{1.25cm}p{1.8cm}p{0.8cm}}
	\hline
	
	\hline
	\textbf{Material} & \textbf{$c_{11}$} & \textbf{$c_{12}$}&\textbf{$c_{44}$}&\textbf{$\overline{M}$}&\textbf{$a$}&\textbf{$K_\mathrm{IC}$}&Ref.\\

	& (GPa) &(GPa) &(GPa) &(GPa) &($\mathrm{\AA}$)&(MPa$\sqrt{m}$)&\\
	\hline
	TiN & 575 & 130 & 163 & 646 & 4.25 &2.05&\cite{Buchinger2019},\cite{Lofler}\\
	CrN & 516 & 115 & 116 & 580 & 4.14 &1.84&\cite{Zhou2013},\cite{Lofler}\\
	\hline
	MgO(100) & 297 & 95 & 155 & 331 & 4.21 &-&\cite{Zha2000}\\
	\hline
\end{tabular}
\end{threeparttable}
\label{tab:elast_prop}
\end{table*}

\section{Results and Discussion}
\subsection{Single layer dislocation density}

\begin{figure}[htb]
\centering
\includegraphics[scale=0.5]{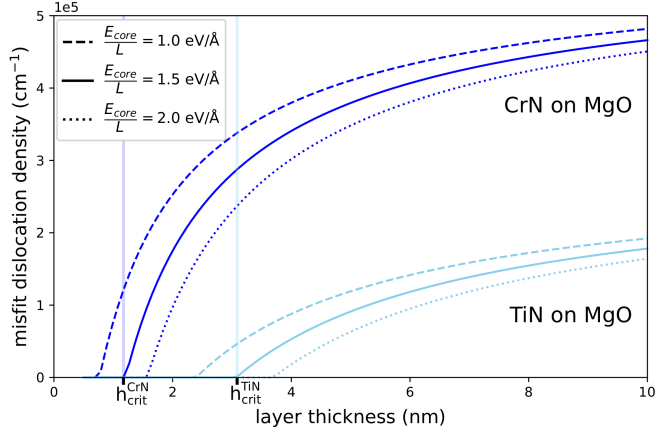}
\caption[Dislocation density]{Misfit dislocation density as a function of film thickness for a TiN and CrN single layer on MgO (100) for different core energies.}
\label{pic:disloc_density_single_layer}
\end{figure}

\noindent Fig.~\ref{pic:disloc_density_single_layer} shows the misfit dislocation densities evaluated for a TiN and CrN thin film single layer, respectively, on an MgO(100) substrate as a function of the film thickness ${h_\mathrm{f}}$. Considering ${\frac{E_\mathrm{core}}{L}}=\SI{1.5}{eV/\AA}$, a critical thickness of $\sim$~\SI{1.2}{nm} for CrN and $\sim$~\SI{3.1}{nm} for TiN is determined. TiN, having a greater lattice parameter than MgO, is in a compressed state when deposited and will relax by a reduction of lattice planes. CrN, on the other hand, is under high tensile coherency stresses, which are reduced by introducing "extra" lattice planes within the film, when the thickness exceeds the critical thickness.\\

\subsection{Multilayer dislocation density}
\begin{figure*}[h]
\centering
\includegraphics[scale=0.42]{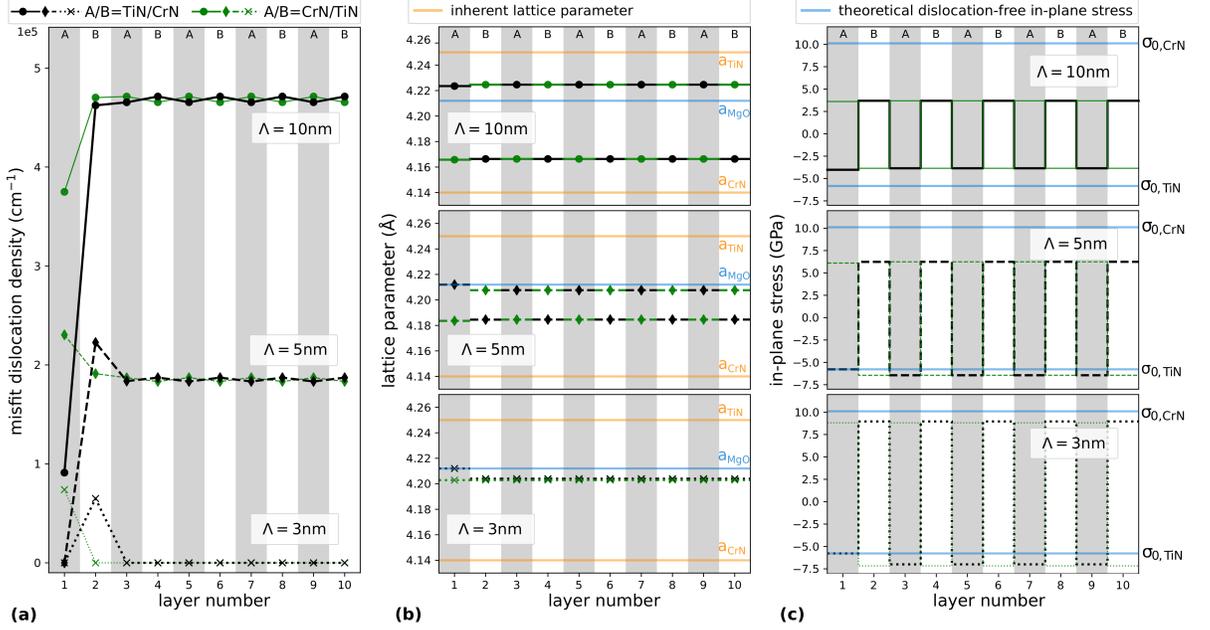}
\caption[Dislocation density]{(a)Misfit dislocation density, (b) resulting strained lattice parameter and (c) in-plane stresses in the individual layers for a superlattice composed of TiN and CrN layers on MgO(100) with three different bilayer periods.
}
\label{pic:disloc_density_stress_TiNCrN_MgO}
\end{figure*}

In the following, we discuss TiN/CrN superlattice films with equally thick TiN and CrN layers, but different bilayer periods ($h_\mathrm{TiN}=h_\mathrm{CrN}=\frac{\Lambda}{2}$) on MgO (100). We choose three different bilayer periods, two corresponding to a layer thickness ${h_\mathrm{l1}}$=\SI{1.5}{nm}=${\frac{\Lambda_1}{2}}$ and ${h_\mathrm{l2}}$=\SI{2.5}{nm}=${\frac{\Lambda_2}{2}}$, where misfit dislocations are predicted for a monolithic CrN film, but not yet for a TiN film. Increasing the bilayer period to ${\Lambda_3}$ with ${\frac{\Lambda_3}{2}}$=${h_\mathrm{l3}}$=\SI{5.0}{nm}, dislocations are expected to form in all layers, regardless of the layer material. We study both situations, starting either with TiN or CrN, subsequently referred to as A/B=TiN/CrN (black curves) and A/B=CrN/TiN (green curves), respectively. Figure~\ref{pic:disloc_density_stress_TiNCrN_MgO}a depicts the resulting dislocation density in the individual layers for the three different architectures for both deposition sequences. For the smallest bilayer period ${\Lambda_1}$=\SI{3}{nm} no dislocations are predicted in the first layer for A/B=TiN/CrN. Hence, the TiN layer adapts to the lattice parameter of the MgO substrate resulting in a strain solely defined by the lattice misfit strain ${\frac{a_\mathrm{MgO}-a_\mathrm{TiN}}{a_\mathrm{TiN}}}$ and the induced curvature. For the subsequently deposited CrN layer a dislocation density of ${Q\approx}$  \SI{6.5e4}{{cm}^{-1}} is energetically favoured. This results in a slight change of the mean lattice parameter towards that of stress-free CrN, see Fig.~\ref{pic:disloc_density_stress_TiNCrN_MgO}b. Even though this induces a higher misfit strain in the next TiN layer, energy minimization after each time a new layer is added, predicts a rather constant lattice parameter, due to the absence of dislocations in the rest of the film. Consequently, the magnitude of in-plane compressive stress in the TiN layers is increased compared to that of a dislocation-free single layer ${\sigma_\mathrm{0,TiN}}$, whereas the tensile stress in CrN layers is decreased compared to ${\sigma_\mathrm{0,CrN}}$ (Fig.~\ref{pic:disloc_density_stress_TiNCrN_MgO}c). Changing the sequence of layer deposition, i.e. starting with CrN, dislocations are formed in the first layer, whereas the rest is predicted to be dislocation-free. The resulting mean lattice parameter and, thus, stress state coincide with the values determined for the TiN/CrN sequence.\\
When increasing the bilayer period to $\Lambda=\SI{5}{nm}$, the formation of dislocations starts again in the first CrN layer, but now alters the mean lattice parameter to an extent high enough to induce misfit dislocations in the TiN layers. The resulting dislocation density is quite similar for both layer materials, being \SI{18.3e4}{{cm}^{-1}} for TiN layers and \SI{18.7e4}{{cm}^{-1}} for CrN layers. The plot in the middle of Fig.~\ref{pic:disloc_density_stress_TiNCrN_MgO}c indicates, that the magnitude of compressive stresses is still higher compared to the mismatch-stresses of a dislocation-free superlattice.\\
A bilayer period of \SI{10}{nm} provokes misfit dislocations in the first layer, independent of the starting layer material. We estimate a dislocation density of \SI{46.5e4}{{cm}^{-1}} and \SI{47.1e4}{{cm}^{-1}} for TiN and CrN, respectively, for both sequences. The stresses are reduced in each sublayer with respect to the dislocation-free thin film. 

\begin{figure}[htb]
\centering
\includegraphics[scale=0.5]{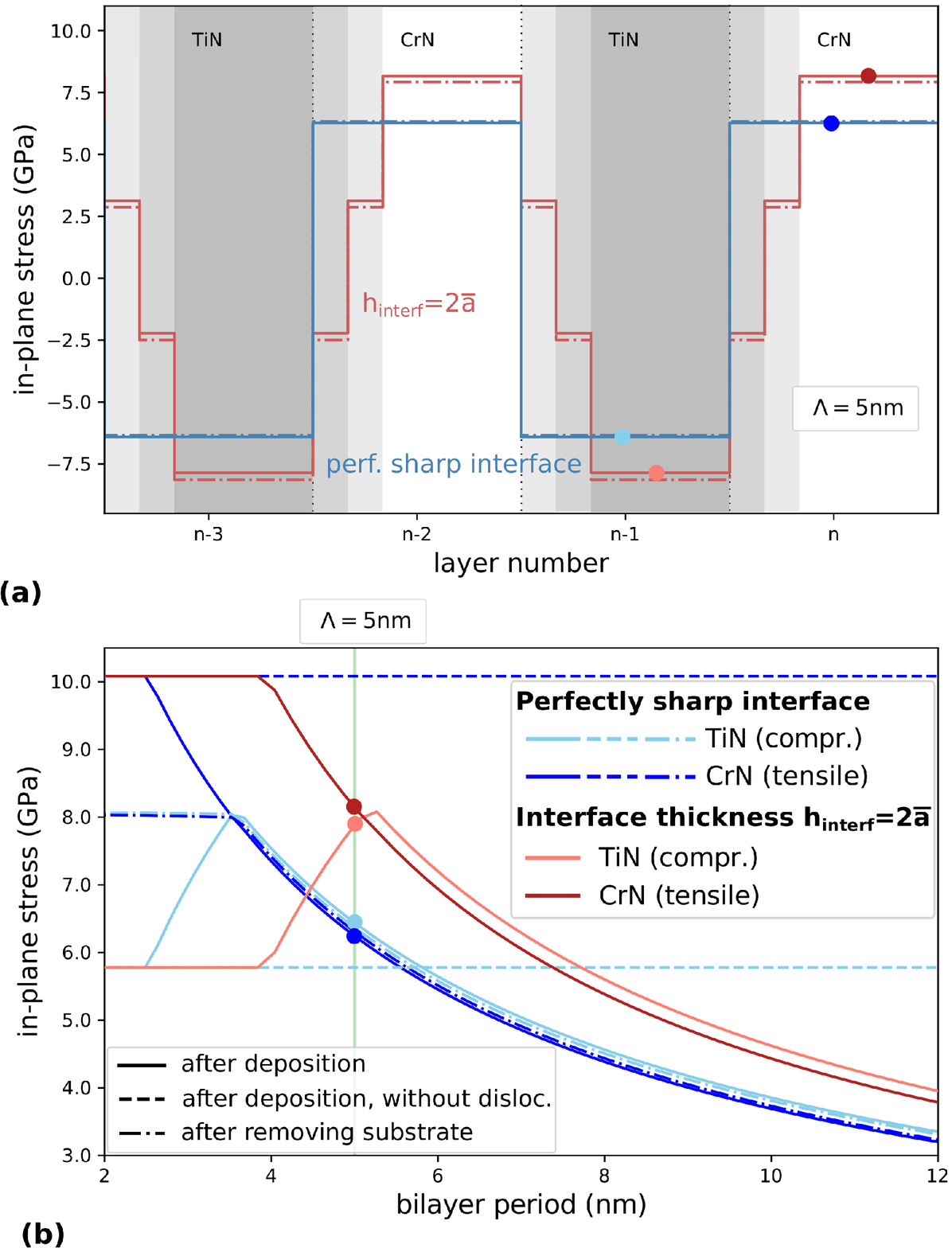}
\caption[Stress vs. bilayer period]{Stress distribution in the top layers of a TiN/CrN superlattice deposited on MgO with total thickness $w=\SI{1.5}{\micro m}$ and $\Lambda=\SI{5}{nm}$ considering either a perfectly sharp interface or an interface thickness of ${h_\mathrm{interf}=2\overline{a}}$ with stepwise composition modulation (a). The dots represent absolute values of stresses plotted as a function of the bilayer period in (b).}
\label{pic:stress_TiNCrN_MgO}
\end{figure}

\noindent For all three architectures, our model predicts that the stress state in the first bilayer slightly deviates from the rather constant stress amplitudes in the rest of the multilayer film. Figure~\ref{pic:stress_TiNCrN_MgO}a exemplarily depicts the stress state of the top four layers of a TiN/CrN superlattice deposited on MgO with $\Lambda=\SI{5}{nm}$ and a total film thickness of $h_\mathrm{f}=\SI{1.5}{\micro m}$. The blue curves correspond to a superlattice film with perfectly sharp interfaces, while the stress distribution in a superlattice film with at finite interface thickness (here $h_\mathrm{interf}=2\overline{a}\approx\SI{8.4}{\AA}$) is shown in red. The dots indicate the magnitude of stress considered in Fig.~\ref{pic:stress_TiNCrN_MgO}b, which presents the bilayer-period-dependency of the stress state. It should be noted that the stress state of both layers is represented by a positive value. The actual type of stresses, compressive or tensile, is indicated within the legend.

For very small bilayer periods, the stress state in both superlattice constituents correspond to their lattice mismatch with respect to MgO (see identical in-plane stress values of solid and dashed lines in Fig.~\ref{pic:stress_TiNCrN_MgO}b). Considering a perfectly sharp interface and increasing the bilayer period to $\Lambda\sim\SI{2.5}{nm}$, dislocations start to form in the CrN layers resulting in a relaxation of tensile stresses (continuous dark blue curve), whereas the compressive stresses in TiN layers (continuous light blue curve) increase until dislocations are introduced within TiN layers as well at $\Lambda\sim\SI{3.6}{nm}$. Accordingly, the curve of compressive stresses shows an increase and subsequent decrease with its peak at $\Lambda\sim\SI{3.6}{nm}$.\\
The results for the bilayer-period-dependency of the stress state with and without finite interface thickness cannot be compared quantitatively, since the stress state within the interfaces is not captured by the values plotted in Fig.~\ref{pic:stress_TiNCrN_MgO}b. However, it can be deduced that a thicker interface shifts the peak of compressive stresses to higher bilayer periods. This is because dislocations start to form at an increased layer thickness compared to the perfectly sharp interface configuration since part of the layer thickness is occupied by the interface with a lower inherent lattice mismatch.\\
As mentioned in Sec.~\ref{sec:method}, evaluating the apparent fracture toughness according to the experimental set-up requires modelling free-standing film material. Removing the substrate induces a stress redistribution within the layers in such a way that, due to the equilibrium conditions, the resulting compressive stresses in TiN are of quite the same magnitude as tensile stresses in CrN for all bilayer periods, see dashed/dotted lines in Fig.~\ref{pic:stress_TiNCrN_MgO}. Hence, the observed peak in compressive stresses is not perceivable anymore after removing the substrate.\\
\subsection{Apparent fracture toughness of multilayers}
\begin{figure}[htb]
\centering
\includegraphics[scale=0.48]{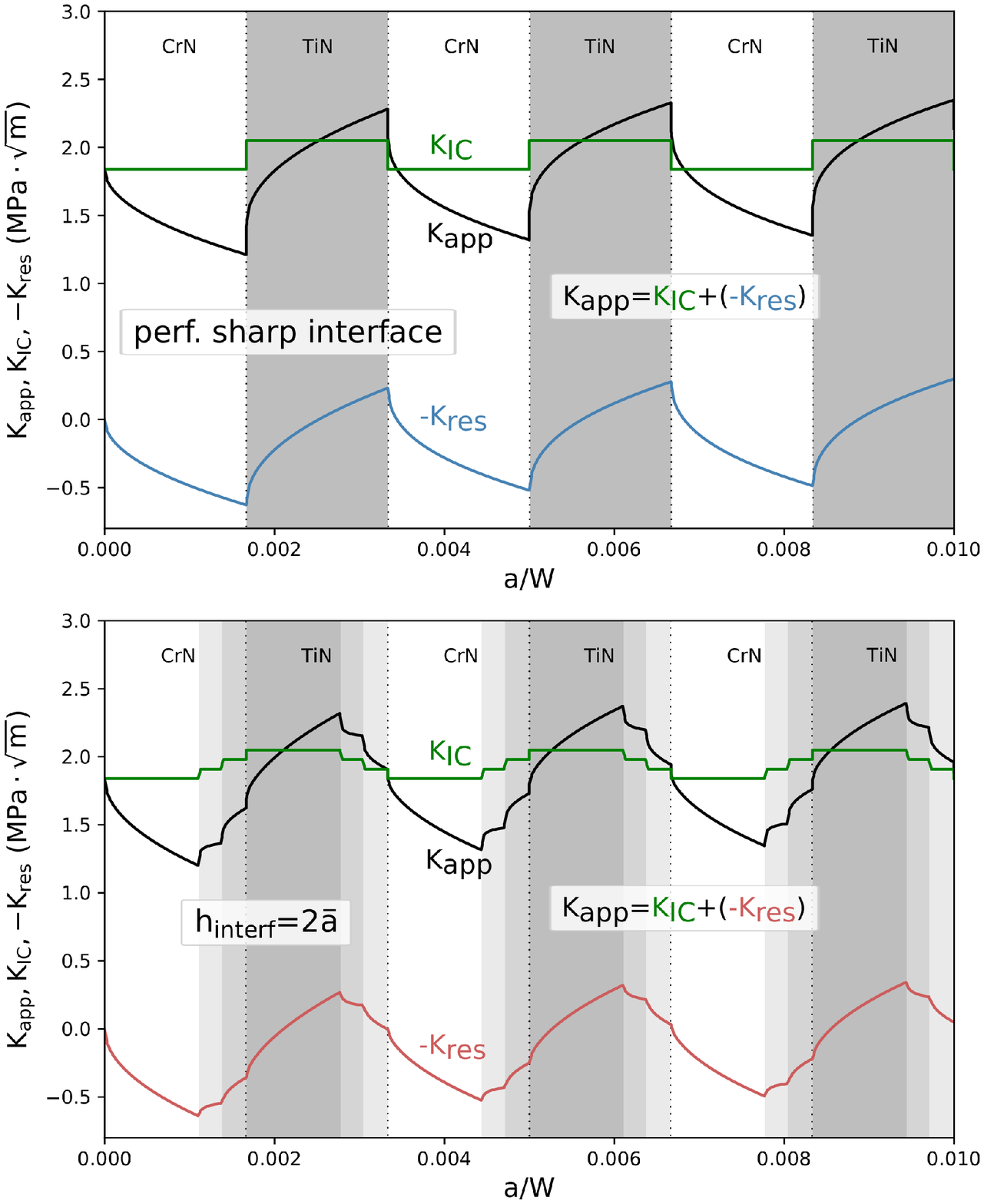}
\caption[Fracture toughness top layers]{Alteration of the apparent fracture toughness in the top few layers of an SL with $\Lambda=\SI{5}{nm}$ resulting from the spatially varying inherent fracture toughness and the stress intensity factor associated with the residual stress state. Top and bottom panels correspond to perfectly sharp interfaces and interface thicknesses of ${h_\mathrm{interf}=2\overline{a}}$, respectively.}
\label{pic:K_app_5nm_plusinterf}
\end{figure}

\begin{figure}[htb]
\centering
\includegraphics[scale=0.495]{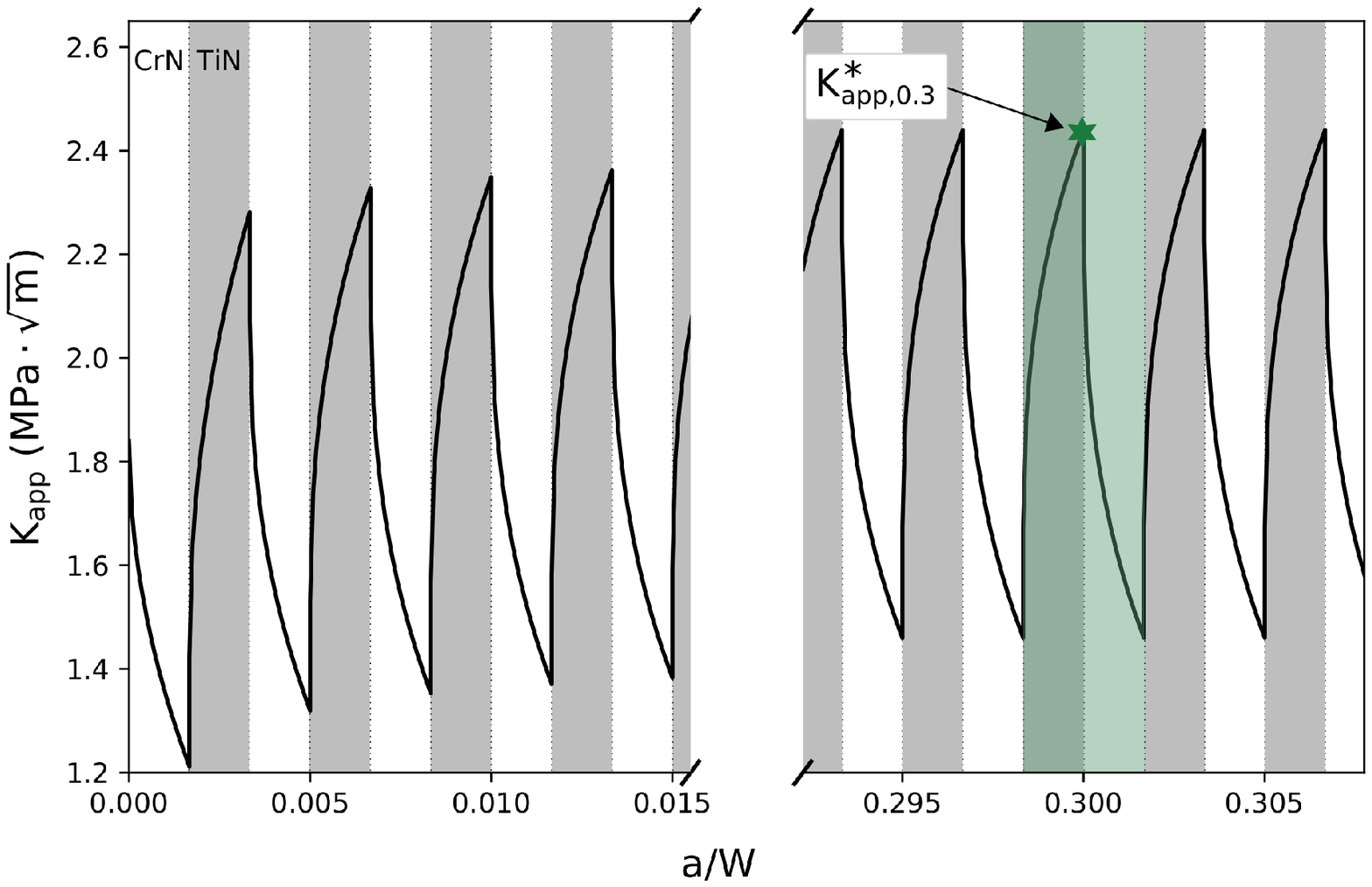}
\caption[Apparent fracture toughness approx. initial crack length]
{Apparent fracture toughness as a function of the ratio between crack length and cantilever thickness $a/W$ for $\Lambda=\SI{5}{nm}$. ${K^*_\mathrm{app,0.3}}$ indicates the considered apparent fracture toughness estimated for an initial crack length of ${a/W\approx0.3}$.}
\label{pic:K_app_5nm_0to0p3}
\end{figure}

\begin{figure}[htb]
\centering
\includegraphics[scale=0.495]{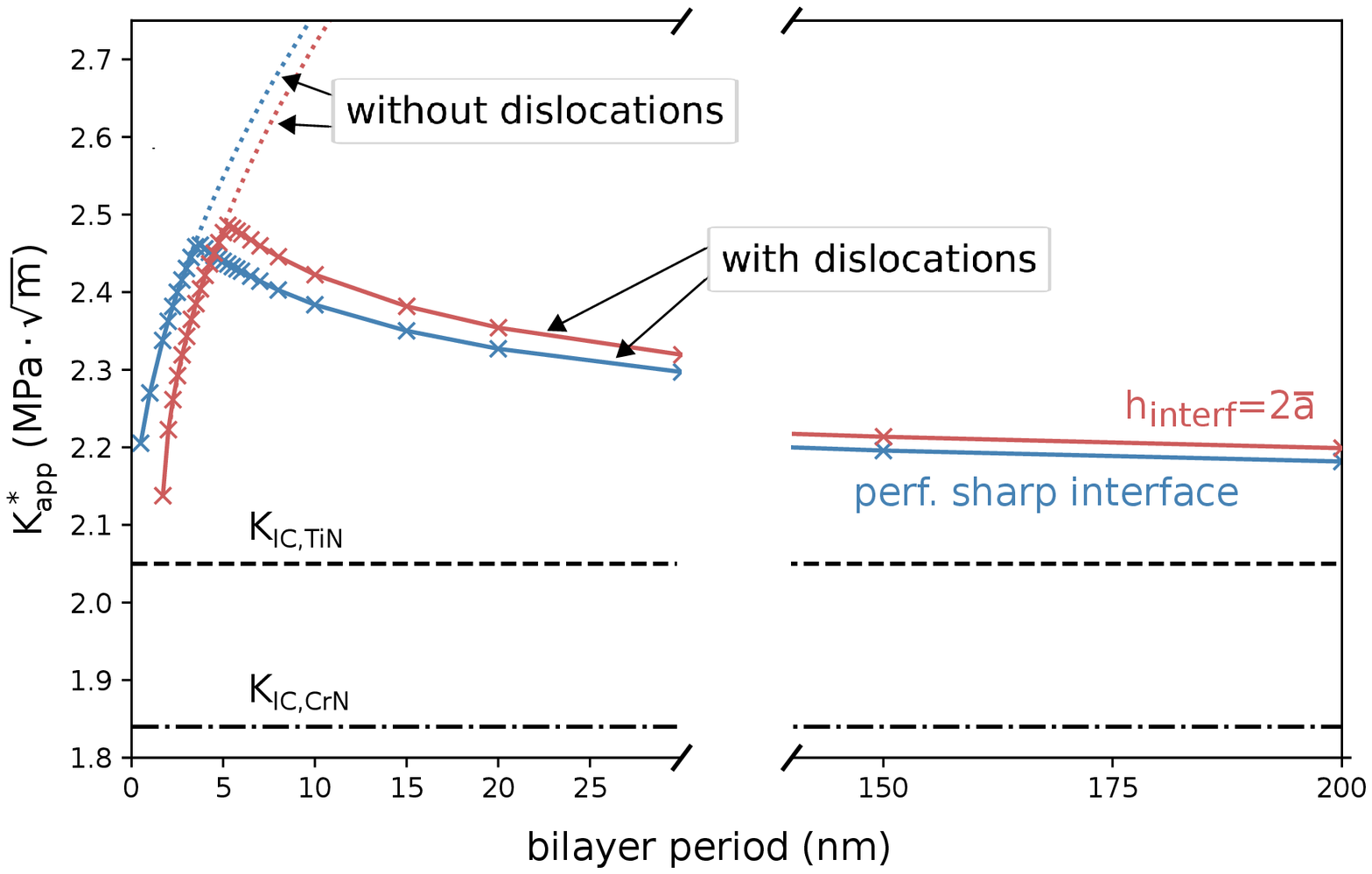}
\caption[Maximum apparent fracture toughness at ${a/W\approx0.3}$ over bilayer period]
{System's apparent fracture toughness estimated for ${a/W\approx0.3}$ as a function of bilayer period for a superlattice with perfectly sharp interface (blue line) and with a finite interface thickness (red line). The dotted line depicts the constant increase of ${K^*_\mathrm{app}}$ if no misfit dislocations were considered.}
\label{pic:K_app_blp}
\end{figure}

For studying the apparent fracture toughness, it should be recalled, that the coordinate system is changed such that its origin is positioned at the free surface of the top layer (corresponding to the right side of Fig.~\ref{pic:stress_TiNCrN_MgO}a). Taking into account the stress distribution depicted by dashed/dotted lines in Fig.~\ref{pic:stress_TiNCrN_MgO}a and applying Eq.~\ref{equ:res_stress_intensity} with the appropriate weight function, we obtain the stress intensity factor resulting from the residual stress state. As discussed in several studies~\cite{Kolednik2005,Kolednik2014}, the crack growth resistance of a multilayer is not only influenced by its stress state but also by the spatial variation of elastic properties of the constituents. However, the influence of the latter is expected to be negligibly small for the considered superlattices, since the constituents are of the same material family with biaxial moduli differing by approximately 10\%.\\Figure~\ref{pic:K_app_5nm_plusinterf} visualizes the spatially varying inherent fracture toughness ${K_\mathrm{IC}}$ (green curves), the stress intensity factor associated with the residual stress state ${K_\mathrm{res}}$ (blue, red curves) and the consequent alteration of ${K_\mathrm{app}}$ (black curves) as a function of the ratio between crack length and cantilever thickness, ${a/W}$. The top three bilayers are depicted. We see that the inherently lower fracture toughness of CrN is further reduced by its tensile stresses, whereas compressive stresses in TiN layers enhance the maximum bearable stress intensity. Consistent with the other material properties, we consider a stepwise modulation of ${K_\mathrm{IC}}$ over the interface thickness (Fig.~\ref{pic:K_app_5nm_plusinterf}b). Despite the higher stresses within the layers when considering an interface thickness of ${h_\mathrm{interf}=2\overline{a}}$ (see the slightly higher peak of the light red curve compared to the light blue curve in Fig.~\ref{pic:stress_TiNCrN_MgO}), there is no significant difference in maximum apparent fracture toughness. This stems from the fact that the layer thickness being subject to these high stresses is reduced by the interface thickness.\\
Fig.~\ref{pic:K_app_5nm_0to0p3} illustrates the behaviour of ${K_\mathrm{app}}$ for a superlattice with $\Lambda=\SI{5}{nm}$ as a function of $a/W$ up to a crack length relevant for fracture toughness experiments on micro-cantilevers. It shows that the maxima, appearing when the crack tip has just penetrated an entire TiN layer, do not vary significantly between consecutive bilayers. Furthermore, it should be noted that one of these maxima will always be reached. A crack with its tip lying within a tensile (CrN-) layer will demonstrate unstable crack growth until reaching a compressive layer. Then stable crack growth will occur until reaching the next local maximum.
We define ${a/W\approx0.3}$ as the initial crack length and consider the maximum value of the apparent fracture toughness within the adjacent layers as the system's apparent fracture toughness ${K^*_\mathrm{app}}$. Presenting this value as a function of the bilayer period, see Fig.~\ref{pic:K_app_blp}, clearly demonstrates an enhancement of ${K^*_\mathrm{app}}$ with respect to the inherent fracture toughness of the superlattice's constituents. Moreover, we predict an initial steep rise and more gentle decrease of ${K^*_\mathrm{app}}$.This behaviour corresponds well with experimentally observed trends~\cite{Hahn2016a}, albeit the simulation data show a slightly less pronounced peak. 
For low bilayer periods, the system's apparent fracture toughness follows the anticipated curve of a system with suppressed dislocation formation. This fictious scenario corresponds to a stress state barely depending on the bilayer period (merely substrate bending alters the stresses to a negligibly small extent). Hence, the constant increase of ${K^*_\mathrm{app}}$ solely stems from the change of individual layer thickness contributing to ${K_\mathrm{res}}$. Even though dislocations in CrN start forming at a smaller bilayer period, the curves match up to $\Lambda\approx\SI{3.6}{nm}$, i.e. the critical bilayer period for the formation of dislocations in TiN. This is because the equilibrium stress state after removing the substrate remains rather unchanged until reaching a bilayer period where dislocations are formed in both layer materials, compare Fig.~\ref{pic:stress_TiNCrN_MgO}b. Similar to the bilayer-period-dependent stress state, also the peak in ${K^*_\mathrm{app}}$ is shifted to higher bilayer periods and reaches a slightly higher maximum when considering an interface thickness of $\mathrm{h=2\overline{a}}$.\\
Decreasing the minimum bilayer period to very low values (lower values than shown in Fig.~\ref{pic:K_app_blp}), presumably would lead to a loss of the layer structure (cf. Refs. \cite{Helmersson1987,Chu1995}). The fracture toughness of the resulting "solid solution" is expected to follow the rule of mixture (hence, further reduction of ${K^*_\mathrm{app}}$ with decreasing bilayer period).

\section{Summary and Conclusions}
\begin{figure}[htb]
\centering
\includegraphics[scale=0.3]{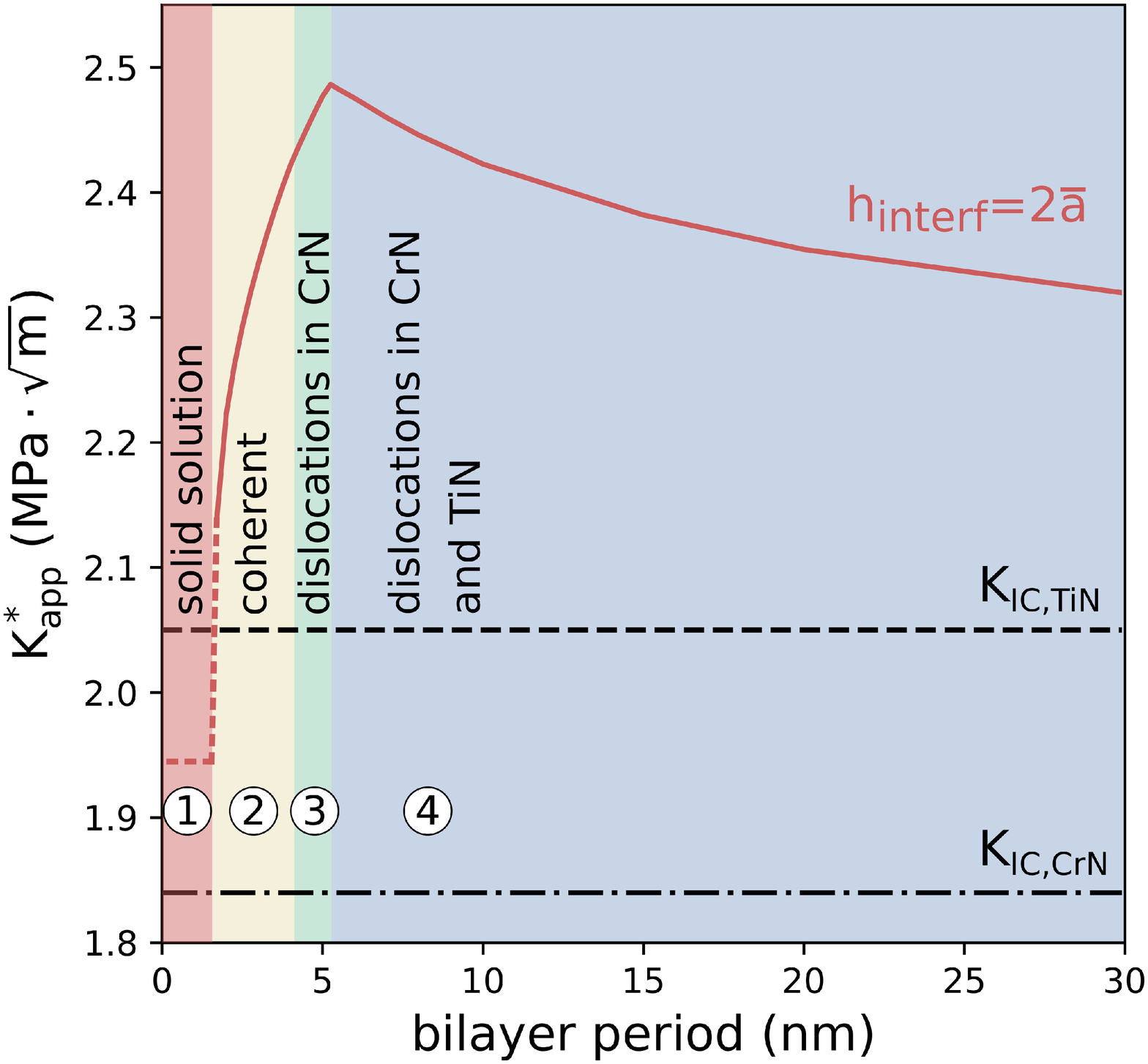}
\caption[Schematic illustration of the different mechanisms influencing the apparent fracture toughness of a TiN/CrN superlattice]
{Schematic illustration of the different mechanisms influencing the apparent fracture toughness of a TiN/CrN superlattice. The dotted line represents the expected behaviour for very low bilayer periods resulting in a "solid solution".}
\label{pic:K_app_conclusion}
\end{figure}

Encouraged by the fracture toughness enhancement experimentally observed in superlattice coatings, we developed a continuum mechanics based model with the main objective to elucidate the underlying mechanisms. By minimizing the overall elastic energy of the substrate/film system each time a new layer is added onto the multilayer stack, we determined misfit dislocation densities as well as evolving coherency stresses for different SL architectures. To allow for comparison with fracture toughness experiments performed on free-standing film material, the substrate was removed after the aforementioned simulation procedure. In contrast to monolithic films, where removing the substrate results in a stress-free state, the stresses in a superlattice are just redistributed within the film. In a final step, we applied the weight function method to link the predicted stress profiles with the crack growth resistance. We found higher critical stress intensity values for all SLs in comparison with the intrinsic fracture toughness of the constituent layer, see Fig.~\ref{pic:K_app_conclusion} (solid line). The crack growth resistance increases with increasing bilayer period for dislocation-free (very thin) SLs (region 2). First dislocations forming in one SL constituent reduce stresses in the corresponding layer material, whereas the strain in the other constituent is increased. However, finding a new equilibrium after removing the substrate results in stresses similar to the dislocation-free configurations, leading to a further increase in apparent fracture toughness (region 3). Only when a critical layer thickness is exceeded and formation of misfit dislocations becomes energetically favorable in both SL constituents, the fracture toughness values decrease again (region 4).

\label{}

\appendix
\section*{Acknowledgement}
The authors highly acknowledge the financial support of the Austrian Science Fund (FWF): P30341-N36.
\appendix
\label{App}

\bibliographystyle{unsrt}
\bibliographystyle{elsarticle-harv}
\bibliography{Paper_superlattice}

\end{document}